# Self-Diffusion and Collective Diffusion of Charged colloids Studied by Dynamic Light Scattering

*Jacqueline Appell#\*, Grégoire Porte#, Eric Buhler#§*

\# Laboratoire des Colloides, Verres, Nanomateriaux (LCVN), UMR5587 CNRS-Université Montpellier II C.C.26, F-34095 Montpellier Cedex 05, France

§ Present Adress: Laboratoire Spectrométrie Physique (LSP), UMR 5588 CNRS-Université Joseph Fourier de Grenoble, BP87, 38402 St Martin d'Hères, France

Jacqueline.Appell@lcvn.univ-montp2.fr    ,    Gregoire.Porte@lcvn.univ-montp2.fr    ,
ebuhler@spectro.ujf-grenoble.fr

\*to whom correspondence should be adressed.

# Abstract

A microemulsion of decane droplets stabilized by a non-ionic surfactant film is progressively charged by substitution of a non-ionic surfactant molecule by a cationic surfactant. We check that the microemulsion droplets remain identical within the explored range of volume fraction (0.02 to 0.18) and of the number of charge per droplets (0 to 40) . We probe the dynamics of these microemulsions by dynamic light scattering. Despite the similar structure of the uncharged and charged microemulsions the dynamics are very different . In the neutral microemulsion the fluctuations of polarization relax, as is well known, via the collective diffusion of the droplets. In the charged microemulsions, two modes of relaxation are observed. The fast one is ascribed classically to the collective diffusion of the charged droplets coupled to the diffusion of the counterions. The slow one has, to our knowledge, not been observed previously neither in similar microemulsions nor in charged spherical colloids. We show that the slow mode is also diffusive and suggest that its possible origine is the relaxation of local charge fluctuations via local exchange of droplets bearing different number of charges . The diffusion coefficient associated with this mode is then the self diffusion coefficient of the droplets.

# Introduction

Dynamic light scatttering (DLS) explores the relaxation of the fluctuations of polarization from which light scattering originates. For suspensions of particles in a solvent (colloidal dispersions, macromolecules or polymers in solution , microemulsions etc...) these fluctuations are essentially due to the fluctuations of concentration. The DLS spectra is a single exponential for dilute dispersions of one species of particles[1,2] reflecting a single relaxation mode ( Brownian motion of the particules) . However in many different experimental situations the DLS spectra is more complicated. For polydisperse suspensions , the spectra is found to be the sum of two exponentials described theoretically [3,4,5] in terms of the collective and self diffusions respectively . These two modes have been oberved in concentrated colloidal dispersions[6,7], in charged dispersions [8,9] and in water-in-oil droplet microemulsions [5,10]. In bicontinuous microemulsions two relaxation modes have been reported[11] which have a different origin related to the local fluctuations of the topology of the bicontinuous network. In linear polyelectrolytes two modes are again observed [12-14], the origin of the slow mode is still controversial. Recently two relaxation modes have been observed in the dynamics of a globular protein and the authors claim the second mode to be due to diffusion of well-ordered clusters of the globular proteins[15] . And in viscoelastic fluids complex DLS spectra are observed with two or three relaxation modes see ref [16] and references therein. One of this mode is directly related to the transient elasticity of the system. This list is not exhaustive, it illustrates the variety of systems where DLS spectra display two or more relaxation modes the origin of which depends strongly on the type of system studied.

In the present paper, we consider the case of weakly charged colloids. Oil in water microemulsion droplets stabilized by a non-ionic surfactant film can be charged by addition of minute quantities of an ionic surfactant without modifying the size and shape of the droplets. They so provide model systems of weakly charged colloid, the charge of which is easily varied and controlled. Such charged microemulsions have been studied previously by Gradzielski and Hoffman [17] and by Evilevitch et al [18] but these authors report on only one relaxation mode in the DLS spectra of these microemulsions. Another convenient realization of a weakly charged colloid is a solution of globular proteins the charge of which is monitored by the pH. Such a system was studied by Retailleau et al [19] who also observe only one relaxation mode in the DLS spectra.

Here, we investigate the DLS pattern of weakly charged oil in water microemulsion droplets. Using small angle neutron scattering, we first check that the size and shape of the droplets remain constant upon changing their charge by addition of small amounts of an ionic surfactant . Then we present the DLS spectra which clearly exhibit two relaxation modes. We then discuss two possible mechanisms at the origin of the two modes. The mechanism first considered by Pusey et al in the early eighties involves the size polydispersity of the droplets. To our knowledge, the second mechanism has never been considered to date: it rather relies on the charge polydispersity of the droplets. A simplified toy model based on a bidispersed distribution of charges in otherwise identical particles is presented to illustrate the probable relevance of this second mechanism.

# Experimental Section

**Materials**

The neutral surfactants are TX100 and TX35 from Fluka used as received. The ionic surfactant: Cetyl pyridinium chloride (CPCl) from Fluka is purified by successive recristallization in water and in acetone. Decane from Fluka is used as received.

**Samples**

The microemulsions[20] are thermodynamically stable dispersions in water of oil droplets surrounded by a surfactant film: O/W microemulsions. The spontaneous radius of curvature of the surfactant film is adjusted by varying the relative proportion of the surfactant and cosurfactant. All samples are prepared by weight in triply distillated water or deuterated water (for the SANS measurements). In the neutral microemulsion, the amphiphilic film is composed of TX100 (surfactant) and TX35 (co-surfactant) ; the weight ratio of TX35 to TX100 is 0.48 ; decane is added so that the sample is close to the emulsification failure limit that is the limit of oil solubilization, the weight ratio of decane to surfactant is 0.7. It is now well establish that, under these conditions, the droplets of microemulsion have a radius corresponding to the optimum curvature of the surfactant film and a narrow size distribution which remain constant over a broad range of concentration [20-22]. We showed previously that it was also the case for this particular microemulsion [23]. In the charged microemulsions a small part (up to 3% in weight) of ionic surfactants are added to the non-ionic surfactants, that corresponds to a number of 0 to 40 ionic polar heads among ~1600 TX neutral polar heads per droplet.

We checked, as described below, that the size of the microemulsion droplets is not modify by this addition. The samples are characterized by the volume fraction of droplets $\Phi$ and by the mean number of charges per droplet $\bar{p}$ ; $\bar{p}$ can be calculated from the composition of the samples and the mean volume of the droplets deduced from the SANS data (see below) . The parameters to calculate $\Phi$ and $\bar{p}$ are given in Table 1 .

Table 1 The components of the microemulsion.

| Component (abbreviation used in the text) | Molar Mass (dalton) | | Density (g/cm$^3$) | |
|---|---|---|---|---|
| | Total | HC(a) | polar part | HC(a) |
| H$_2$O | 18 | - | 1 | - |
| [H$_3$C-(C-(CH$_3$)$_2$-CH$_2$-C-(CH$_3$)$_2$)φ ](O-CH$_2$-CH$_2$)$_3$ -OH   (TX35) | 338 | 189 | 1.2 | 0.86 |
| [H$_3$C-(C-(CH$_3$)$_2$-CH$_2$-C-(CH$_3$)$_2$)φ](O-CH$_2$-CH$_2$)$_{9.5}$-OH   (TX100) | 624 | 189 | 1.2 | 0.86 |
| [H$_3$C-(CH$_2$)$_{15}$ ] -C$_5$H$_5$N$^+$ Cl$^-$    (CPCl) | 339.5 | 225 | 1.656 | 0.83 |
| [H$_3$C-(CH$_2$)$_8$ CH$_3$]           (decane) | 142 | 142 | - | 0.75 |

(a) HC= Hydrophobic part of the molecule in brackets in the formula on the left

**Small Angle Neutron Scattering : SANS Measurements**

The intensity scattered by a dispersion of spherical colloids with a distribution of size can be written [24]:

$$I(q) = I0(q) S^M(q) \quad \text{and} \quad I0(q) = \Phi v_m^{-1} <A^2> \quad (1)$$

With $q$ (Å$^{-1}$) the scattering vector, $\Phi$ the volume fraction, $v_m$ and $<A^2>$ respectively the volume and the scattered amplitude squared of the colloids averaged over the size distribution. $S^M(q)$, the measurable structure factor , reflects the interactions between the colloids, it is 1 at large q's or for very dilute samples where interactions can be neglected.

The best simulation is obtained for spherical droplets of outer radius $R_3$ but with three successive scattering length densities corresponding roughly to an inner sphere of decane, a corona containing the aliphatic chains of surfactant and an outer corona containing the polar heads = short PEO chains . The scattering length density profile is sketched in figure 1 . The averaged squared amplitude is calculated from the amplitude for spherical concentric shells[25] and with a Schulz-Zimm distribution for the size of the droplets[26]

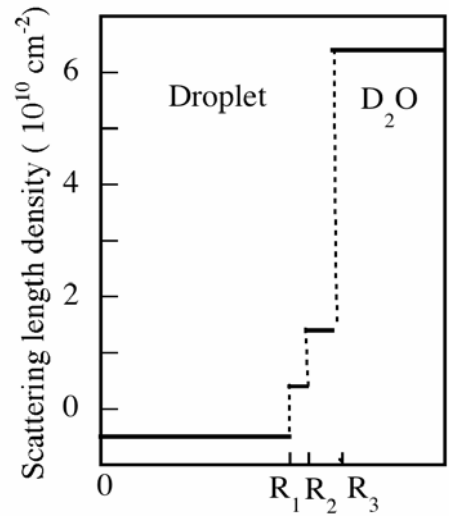

Figure 1 : Scattering length density profile.

The best fit is for $R_1 = 6.6$ nm , $R_2 = 7.2$ nm and $R_3 = R_m = 8.1$ nm with a standard deviation = 1.5 nm ( the two corona retain constant width) .

The experiments have been performed at LLB-Saclay on the spectrometer PACE where the range of scattering vectors covered is 0.006 Å$^{-1}$ < $q$ < 0.16 Å$^{-1}$ . The temperature is T= 22°C. The scattering data are treated according to standard procedures. They are put on an absolute scale by using water as standard. We thus obtain intensities in absolute units (cm$^{-1}$) with an accuracy better than 10%. To simulate correctly the experimental spectra, the model spectra are convoluted with the instrumental response function taking into account the actual distribution of the neutrons wavelength and the angular definition [27] . In Figure 2, the SANS spectra for samples with Φ = 0.038 are shown. The repulsive interactions between droplets increase with increasing charge as evidenced in figure 2A , these interactions will be examined in a forthcoming paper. The form factor oscillations -damped by the size distribution- are amplified in the $q^4 \times I(q)$ representation in figure 2B . These oscillations are identical for all the samples for q > 0.04 (when $S^M(q) \approx 1$) and they are reasonably well reproduced with $I0(q)$ calculated as indicated above . The microemulsion is well represented by a suspension of spheres ( $R_m = 8.1 \pm 0.2$ nm, $\Delta R = 1.5$ nm , $v_m = (2.5 \pm 0.3) \times 10^{-18} cm^3$ ) with a small size polydispersity ( $p = \frac{\Delta R}{R_m} = 0.18$ ) and we checked, that the microemulsion droplets remain identical at least for $0.02 < \Phi < 0.18$ and $0 < \bar{p} < 40$.

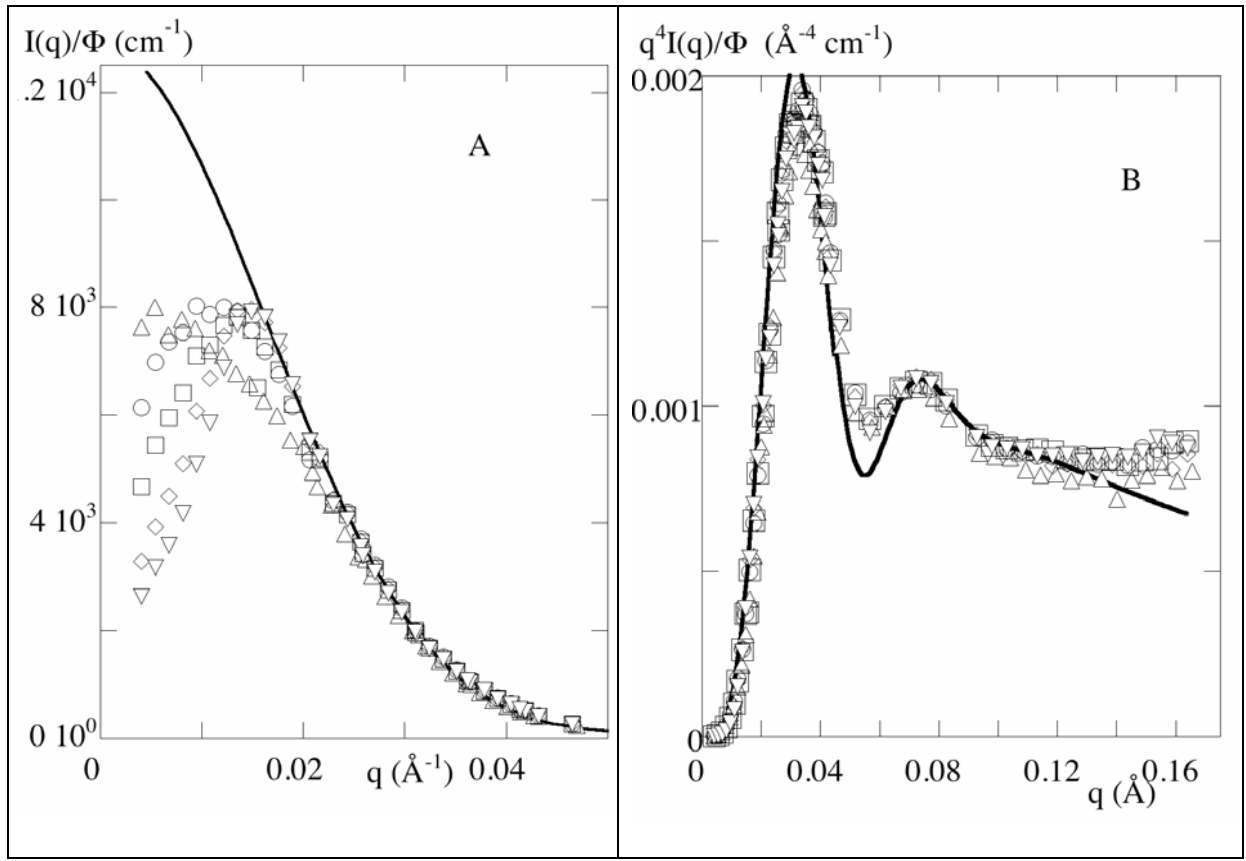

Figure 2 SANS Spectra for the microemulsions with $\Phi = 0.038$. The line corresponds to I0(q) computed for spherical droplets with the scattering length density profile shown in fig1 with an outer mean radius $R_m = R_3 = 8.1\ nm$ and a standard deviation $\Delta R = 1.5\ nm$. Triangles: $\bar{p}$ =0; circles: $\bar{p}$ =5.5; squares : $\bar{p}$ =11; diamonds: $\bar{p}$=16.5; inverse triangles: $\bar{p}$=22. Fig.2A: I(q) as a function of q illustrates the increasing repulsive interaction brought by the charges on the droplets. Fig 2B the Porod representation $q^4 I(q)$ as a function of q emphasizes the form factor oscillations which are identical in the high q range: the drops are unchanged in the charged microemulsion.

## Dynamic light scattering - DLS

The thermally activated dynamics of microemulsions can be probed by DLS[1,2]. The measurements are performed on a standard setup (AMTEC Goniometer with a BI9400 Brookhaven correlator), the light source is an argon ion laser ($\lambda$=514.5 nm). The homodyne intensity autocorrelation function is measured at different q values, ranging from 3 10$^6$ to 3 10$^7$ m$^{-1}$ ( $q = \frac{4\pi n}{\lambda} \sin\frac{\theta}{2}$ with *n* the refractive index of the solvent and $\theta$ the scattering angle ).

The normalized autocorrelation function of the scattered intensity writes:

$$g_2(q,t) = \frac{\langle I(q,t) I(q,0) \rangle}{|\langle I(q,0) \rangle|^2} \qquad (2)$$

and if the scattered field obeys Gaussian statistics, this measured autocorrelation function can be expressed as a function of the first-order electric field correlation function through the Siegert relationship: $g_2(q,t) - 1 = c \left| g_1^E(q,t) \right|^2$, where c (0<c<1) is the experimental coherence factor. In our samples, the scattering originates from the fluctuations of the droplet concentration and $g_1^E(q,t) \propto g_1(q,t)$ the autocorrelation function of the concentration fluctuations. In a suspension of microemulsion droplets in the low concentration regime we expect these fluctuations of concentration will relax through simple brownian diffusion and, assuming a monodisperse suspension, we simply write :

$$g_1(q,t) = \exp\left[-\frac{t}{\tau}\right] \text{ with } \tau^{-1} = D_{coll} q^2$$

where $D_{coll}$ is the usual collective translational diffusion coefficient.

So that $g_2(q,t) - 1 = c \exp\left[-\frac{2 \times t}{\tau}\right]$  (3)

In this case the relaxation mode is diffusive as indicated by the $q^{-2}$-dependence of $\tau$. This is indeed what is observed in the neutral microemulsions as illustrated in figure 3A, the small departure from a monoexponential behavior at large times is to trace back to the moderate size polydispersity of the droplets. The well-known cumulants expansion [4,28] -strictly valid for a non interacting suspension- can be used to account for the polydipersity of the microemulsion (cf fig 3A).

$$Ln(g_1(q,t)) = b - \overline{\Gamma}t + \frac{var}{2}(\overline{\Gamma}t)^2 + ...$$

with $\overline{\Gamma} = \overline{\tau}^{-1} = \int \Gamma G(\Gamma) d\Gamma$ and $var = \frac{\int (\Gamma - \overline{\Gamma})^2 G(\Gamma) d\Gamma}{\overline{\Gamma}^2}$ (4)

where $G(\Gamma)$ is the distribution of $\Gamma's$

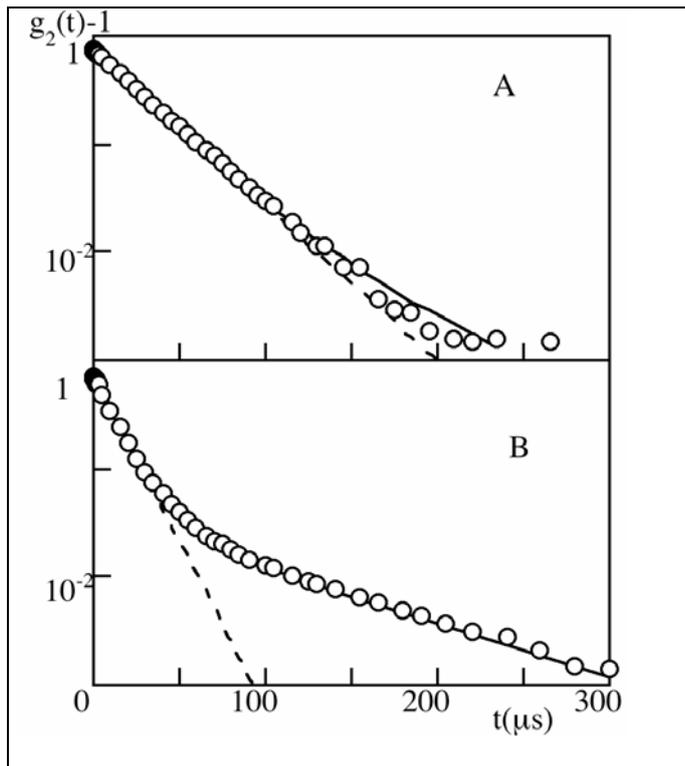

Figure 3 : The normalized autocorrelation function measured at $\theta = 90°$ for a neutral microemulsion (A) and a charged microemulsion $\overline{p} = 40$. (B) The volume fraction $\Phi = 0.17$. The points are experimental data; the dotted lines are fits of the data to a single exponential (eq 3). The line in A is a fit to eq 4 with $\overline{\Gamma} = 1.8 \pm 0.01 \, \mu s^{-1}$ var = 0.09. The line in B is a fit to eq (5) with
$\alpha_f = 0.8$, $\tau_f = 20 \pm 2 \mu s$ and
$\alpha_s = 0.2$, $\tau_s = 180 \pm 20 \mu s$.

The situation is different when the microemulsion droplets are charged. As can be seen in figure 3B the autocorrelation function is no longer quasi-monoexponential and it is well fitted to the sum of two single exponentials:

$$g_2(q,t) - 1 = c \left[ A_f \exp\left[-\frac{t}{\tau_f}\right] + A_s \exp\left[-\frac{t}{\tau_s}\right] \right]^2 \quad (5)$$

Two relaxation modes are present, each one characterized by its relative contribution and its decay time : $\alpha_i = \dfrac{A_i}{A_f + A_s}$ and $\tau_i$ $i = f$ (fast) or $s$(slow). This was found to be the case for all the charged microemulsions as is further illustrated in figure 4 for samples with a constant charge $\bar{p}=20$ and an increasing volume fraction.

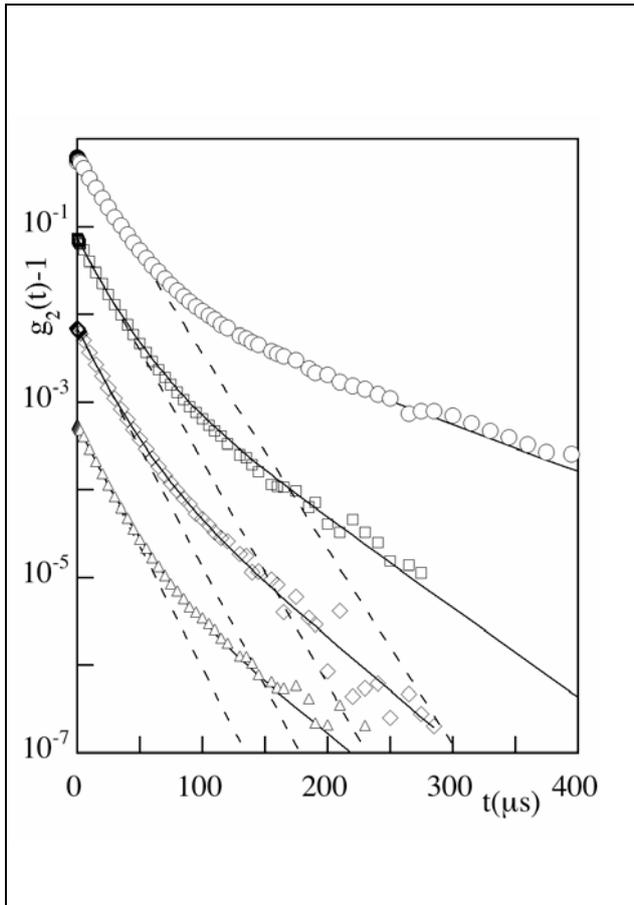

Figure 4 : Evolution with the volume fraction of droplets of the normalized autocorrelation function measured at an angle of 90° for samples with a constant number of charges $\bar{p} = 20$. The lines are fits of the data to eq.(5) and the dotted lines are fits of the data to a single exponential. Each set is shifted by a factor of 10 for clarity. Circles: $\tilde{\Phi}_1$ ; squares: $\tilde{\Phi}$ ; diamonds $\tilde{\Phi}$ 35; triangles : $\tilde{\Phi}$ 17.

The two relaxation modes are diffusif: $\tau_i^{-1} = D_i\, q^2$ as illustrated in figure 5 . The slopes of the straigth lines are the diffusion coefficients $D_f$ and $D_s$ associated to the two relaxation modes .

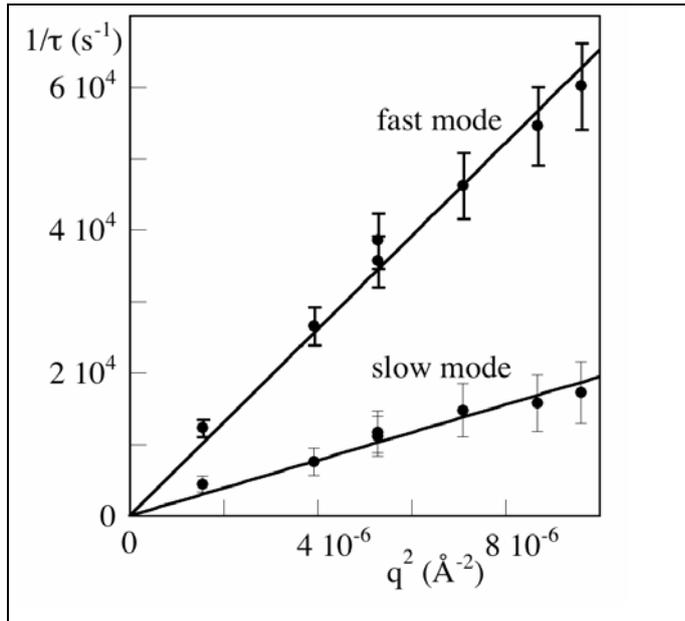

Figure 5 : The two modes are diffusion modes: $1/\tau = D q^2$ : illustration for the sample $\tilde{\Phi}$ , $\bar{p}=20$ . From the slopes of the straight lines we obtain the diffusion coefficients $D_f = (6.5 \pm 0.5) \times 10^{-11} m^2 \times s^{-1}$ and $D_s = (1.9 \pm 0.2) \times 10^{-11} m^2 \times s^{-1}$

Furthermore the relative amplitudes of the two modes are q-independent within experimental errors as shown in figure 6 .

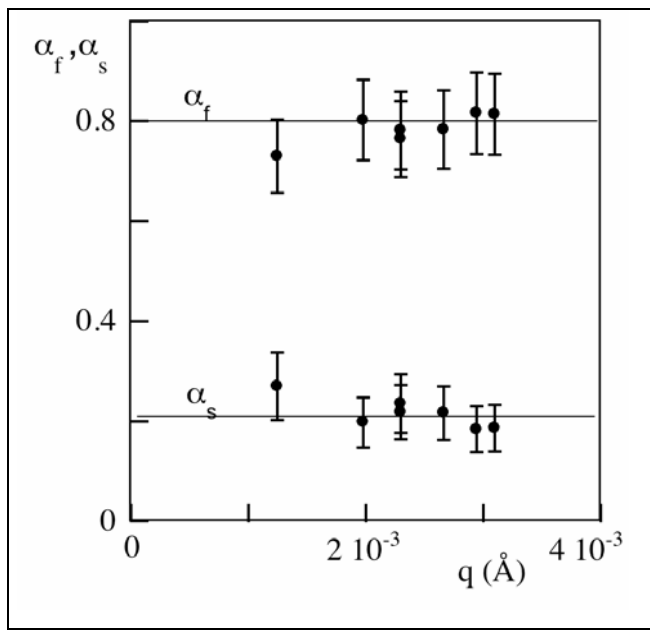

Figure 6 : The relative amplitudes of the two modes do not depend on the wave vector q. Illustration for the sample $\tilde{\Phi}$ , $\bar{p}=20$ .

These features are found for the different sets of samples. We will now summarize the results. We already stressed that both modes of relaxation are diffusive , they are characterized by a diffusion coefficient. The diffusion coefficient varies with the volume fraction and with the number of charges per droplet in a different fashion for both modes. This is illustrated in figures 7 and 8 . The salient features are:

i/ The 'unique' diffusion coefficient for the neutral microemulsion increases weakly with $\Phi$, a behavior generally observe for the collective diffusion coefficient of dilute microemulsions where the droplets interact via an overall repulsive potential (a Van der Waals attractive potential plus a hard sphere repulsive potential) [29]. From $\overline{D}_0 = 2.58 \times 10^{-11} \, m^2 \times s^{-1}$ we can extract the apparent hydrodynamic radius [2] $R_{app} = 95$ of the droplets using the Stokes-Einstein relationship: $\overline{D}_0 = v_0 k_b T = \dfrac{k_b T}{6\pi \eta R_{app}}$ with $k_b$ the Boltzmann constant, $T$ the temperature, $v_0$ the mobility at infinite dilution and $\eta$ the viscosity of the solvent (here water). The value obtained for $R_{app} = 9.5$ nm and the normalized variance of the distribution of diffusion coefficients equal to a few percent, compare well to the values calculated for droplets with a mean geometrical radius of 8.1 nm and a standard deviation of 1.5 nm [8].

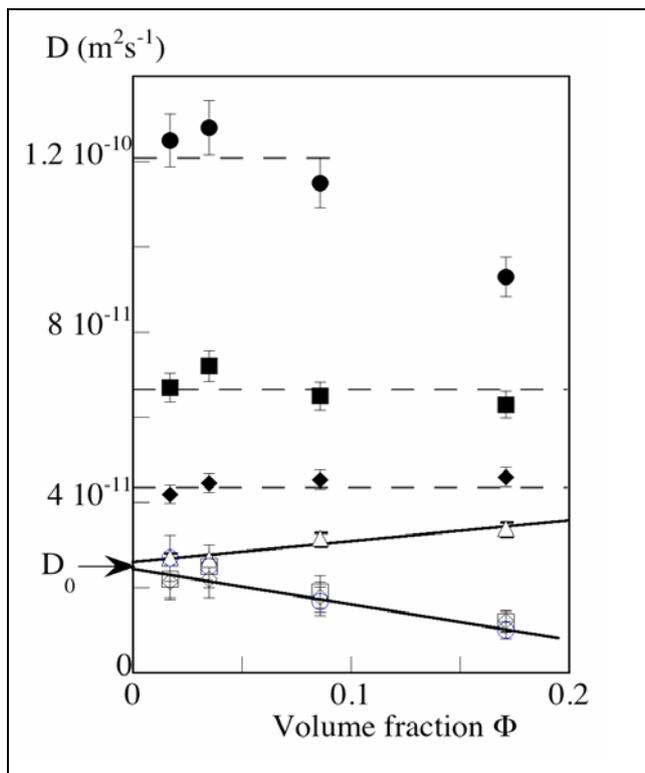

Figure 7 The diffusion coefficients for the two modes as a function of the volume fraction of droplets. The closed symbols are for the fast mode, the dotted lines are guide for the eyes. The open symbols are for the slow mode the line through all the data for the slow mode is a fit to a straight line. Circles $\overline{p}=40$; squares $\overline{p}=20$, diamonds $\overline{p}=10$. For comparison the diffusion coefficients for the single mode in the neutral microemulsion is shown : open triangles and the straight line is a fit to the data. See text for further discussion.

ii/ the diffusion coefficient $D_s$ associated with the slow mode does not depend, within experimental errors, on the mean number of charges per droplet $\overline{p}$ so that a straight line can be drawn through all the points. This line extrapolates to the value $D_0$ at $\Phi = 0$. However, as can be seen in figure 7, $D_s$ decreases with increasing $\Phi$.

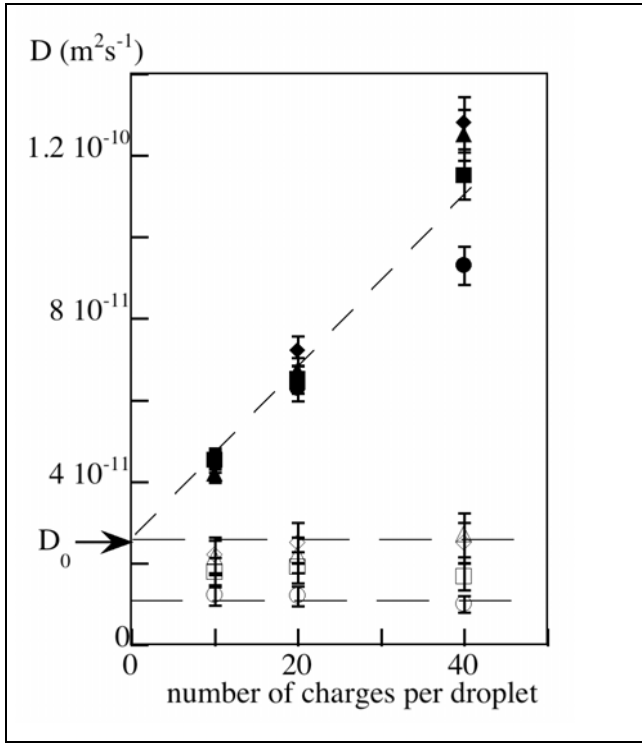

Figure 8 The diffusion coefficients for the two modes as a function of the number of charges per droplet. The closed symbols are for the fast mode, the open symbols are for the slow mode. The dotted lines are guide for the eyes. Circles $\Phi = 0.17$; squares: $\Phi = 0.086$; diamonds: $\Phi = 0.035$; triangles $\Phi = 0.017$.

iii/ The diffusion coefficients $D_f$ associated with the fast mode do not depend on $\Phi$ except possibly at the highest $\bar{p}$ and $\Phi$ but in contrast with the diffusion coefficients of the slow mode they depend heavily on the number of charges. In figure 8 it can be seen that the extrapolation value for the fast mode at zero volume fraction and zero charge is also $D_0$. However the way the diffusion coefficients for both modes extrapolate to $D_0$ is very contrasted.

iv/ A last observation is worth stressing namely that, as can be seen in figure 9, the relative amplitudes of the two modes remain pratically constant within experimental errors for all samples. The slow mode contributes roughly to one fifth of the overall relaxation.

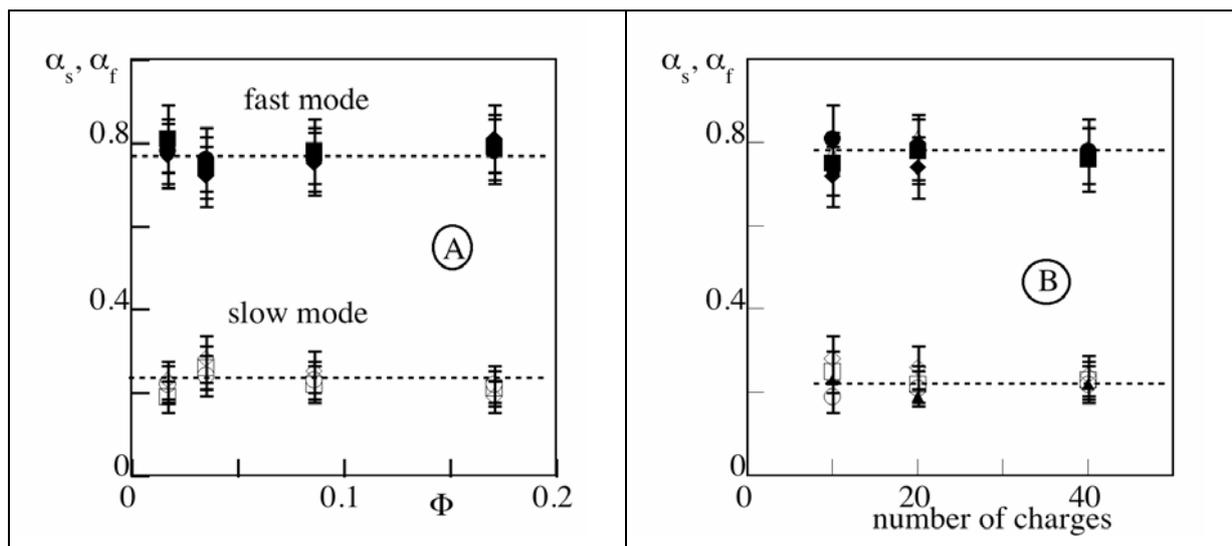

Figure 9 Relative amplitudes of the two modes. A: as a function of volume fraction of droplets Circles $\bar{p}=40$; squares $\bar{p}=20$, diamonds $\bar{p}=10$.and B as a function of the number of charge per droplet. Circles $\Phi=0.17$; squares: $\Phi=0.086$; diamonds : $\Phi=0.035$ ; triangles $\Phi=0.017$. The closed symbols are for the fast mode , the open symbols are for the slow mode. The dotted lines are guides for the eyes.

## Discussion.

We stressed in the introduction that in the microemulsion as in other colloidal suspensions , the fluctuations of polarization which are the cause of light scattering essentially originate from the fluctuations of concentration of the colloids . In the present sample this is also the case , what is then the physical origin of the different behaviors between the neutral microemulsion which shows, as usual, one diffusive mode and the weakly charged microemulsions which clearly show two modes ? Two possible origins can be proposed . One is the size polydispersity of the microemulsion which leads to a distribution of the scattering power of the droplets , it exists in the neutral and charged microemulsion . The other is the possibility that in the charged microemulsions, where on the average each droplet carry $\bar{p}$ charges, this number can fluctuate from droplet to droplet with a distribution at thermal equilibrium . In both cases , the microemulsion droplets are no longer identical so that detection of self diffusion is, in principle, possible . We discuss each one in turn below .

In a colloidal suspension the size polydispersity induces a a distribution of the scattering power of the particles which are thus no longer indiscernable . As described in the pioneering work of Pusey et al[3,4] more than twenty years ago, the autocorrelation function $g_1(q,t)$ splits in two terms one corresponding to the collective diffusion and one to the self diffusion of the colloidal particles . The collective diffusion relaxes a large scale fluctuation of the local number density of the colloids regardless of the local scattering power . The collective diffusion coefficient can be written : $D_{coll} = \frac{\nu k_b T}{S(0) f_c} = \nu \frac{d\Pi}{dN}$ with S(0) the structure factor in the limit q=0, $\nu$ the mobility, $\Pi$ the osmotic pressure and $N$ the number density of particles . In this step the colloids readjust their distance but remain in their initial spatial arrangement (each particle keeping the same nearest neighbours) . Once the number density has relaxed to the homogeneous average value, residual scattering may still persist arising from the local fluctuations of the relative proportion of the big (higher scattering power) versus small (lower scattering power) particles . Such residual contribution to the fluctuation of scattering length density can only relax through the self diffusion of individual colloids and the self diffusion coefficient writes : $D_{self} = \nu k_b T$ . It must be stressed however that the two diffusion coefficients are close to one another as long as the interactions between the colloids are small (they become equal for non interacting colloids) . The two modes are thus detected only in situations where the interactions are large enough so that the two diffusion coefficients become significantly different . This provides a first plausible interpretation of our results :

i/ in the neutral microemulsion the interactions are weak : the two modes cannot be distinguished . ( In neutral colloidal supsensions indeed the two modes are observed only for fairly large volume fractions for which interactions are strong [5-7,10] .)

ii/ in the charged microemulsion the interactions are large as illustrated in figure 2A. The two diffusion coefficients become different ($D_{coll}$ increases and $D_{self}$ decreases) and furthermore the amplitude of the collective mode ($\alpha$ S(0)) is suppressed by the interactions so that the two modes can be readily observed. Two modes have indeed been observed in charged colloidal systems [8,9,30,31].

Although this mechanism based on the size (and therefore the scattering power) polydispersity certainly contributes to our observation, we presume that a second mechanism rather based on the charge polydispersity may well bring a significant contribution.. It could lie in the main difference between the two situations namely that in the charged microemulsions, where on the average each droplet carry $\bar{p}$ charges, this number can fluctuate from droplet to droplet with a distribution at thermal equilibrium. The repulsions between droplets depend strongly on the number of charges they bear: droplets of higher charge are indeed more repulsive. Thus, here again, we must not consider only one fluctuating degree of freedom (namely the local number density of droplets) but we must also consider the local mean density of charges. Considering the relaxation of a large scale concentration fluctuation, we then distinguish two steps. In the first step, the droplets move collectively in order to equilibrate their electro-osmotic repulsions : regions where the initial electro-osmotic repulsion is higher swell at the expense of those where the repulsion is lower. This step involves only rearranging the distance between neighbouring droplets and does not imply any change in the topology of the spatial distribution of the droplets (the neighbours of any given droplet are the same at the end of this step: simple local swelling). Moreover, the force, driving the droplets motion, is the electro-osmotic force. In the present experimental situation, this fast step relaxes about 80% of the fluctuation and it merges into the unique diffusion mode at zero charge. Thus we conclude that this first step is driven by the collective diffusion of the droplets at quenched number of charges per droplet surrounded by their counterions.

We identify the diffusion coefficient arising from the fast mode to the collective diffusion coefficient of the microemulsion.

At the end of this step however, there remains an unreleased contribution to the droplets concentration fluctuations associated with the fluctuations of the mean charge per droplet: regions where the charge per droplet is higher being indeed less concentrated. Thus, to achieve complete equilibrium, we must also ensure that the local mean density of charge has everywhere reached the average equilibrium value. One possibility to equilibrate the charge density relies on the spontaneous thermally activated exchange of ionic surfactants from droplet to droplet. However, we know from previous investigations (rheometry)[16,32] on similar droplet dispersions linked together by telechelic polymers that the residence time of the stickers of comparable aliphatic chain length in the droplets is of the order of a few $10^{-2} s$. This is much longer than the relaxation of the intensity–intensity correlation function.

So we conclude that the number of charges on each droplet remains frozen during the time of the experiment so the mechanism which dominates the equilibration of the charge density fluctuations is the relative diffusions of high versus low charge droplets. In this picture, the slow mode simply corresponds to the self diffusion of droplets and the associated diffusion coefficient is the self-diffusion coefficient.

Assuming this description of the two relaxation modes, we can now examine if the experimental results sustain it. $D_s = D_{self}$, then $D_s = \nu\, k_B T$  $\nu$ is the mobility which depends on the hydrodynamics of the microemulsion, $D_s$ depends on $\Phi$ but not on $\bar{p}$, this implies that the charges have no or little influence on the mobility of the droplets and $D_s$ extrapolates to $D_0$ at zero volume fraction as it should. On the other hand $D_f = D_{coll}$ is determined by the osmotic compressibility and the mobility of the droplets. Therefore we expect it to depend strongly on the charge per droplet (fig 8). On the other hand (fig.7), at a given $\bar{p}$ it does not depend (or weakly) on $\Phi$, in the range explored, probably because of subtle compensation between the osmotic compressibility and the mobility of the droplets. Anyway we do not observe, as expected in the dilute regime,, $D_f = D_{coll} = D_0$. This can be understood if we recall that the electrostatic potentiel of interaction is long-ranged (unscreened coulombic potential) so that the dilute regime (in the sense that the interactions between droplets become negligible) is much lower than the lowest limit studied ($\Phi = 0.017$) and extrapolation to zero volume fraction at a finite $\bar{p}$ cannot be achieved properly. However, $D_f$ extrapolates to $D_0$ at zero volume fraction and zero charge as it should. We find no contradiction between our description and the evolution of the two diffusion coefficients.

To illustrate our general description, let us consider the very simple model described in the next section.

## A simple model for the diffusion of charged particles

Although the charge distribution of the droplets is indeed spread, we consider for simplicity two populations of particles only which differ by their number of charges per particles $p_1$ and $p_2$. Their respective concentrations (in molar fractions) are noted $X_1$ and $X_2$. All particles are otherwise identical (same size, shape and scattering length). We assume further that the free energy density $f$ (in molecular units) of the dispersion comprises two terms:

$$f = h(p_1 X_1 + p_2 X_2) + k_b T (X_1 \log_e X_1 + X_2 \log_e X_2) \tag{6}$$

The first term $h$ is the enthalpic contribution arising from the repulsion between the particles due to their charges. Its functional form is, but at the scale of the inverse wave vector investigated experimentally (large scale compared to the average distance between the droplets), we can reasonably admit that it is a function of the local charge density of the dispersion only. The second term is indeed the usual contribution of the entropy of dilution of populations 1 and 2 respectively. Expanding $f$ in terms of small concentration fluctuations $\delta X_1$ and $\delta X_2$ from the average values $\bar{X}_1$ and $\bar{X}_2$:

$$f = f_0 + \text{linear terms in the } \delta X_i's + \frac{1}{2}\left(p_1^2 A + \frac{k_b T}{\bar{X}_1}\right)\delta X_1^2 + \frac{1}{2}\left(p_2^2 A + \frac{k_b T}{\bar{X}_2}\right)\delta X_2^2 \qquad (7)$$
$$+ p_1 p_2 A \, \delta X_1 \delta X_2 + \text{ higher order terms}$$

where $A$ stands for the second derivative of the enthalpic repulsion with respect to the local charge density:

$$A = \frac{\partial^2 h}{\partial (p_1 X_1 + p_2 X_2)^2}$$

Of course only the quadratic terms in (eq.7) are relevant for the diffusion of the particles. For the sake of simplicity again, we further assume that the hydrodynamic interactions between the two populations can be neglected. This is a strong approximation, but we believe it not to be too dramatic at sufficient dilution. In this limit, we can therefore use the generalized Fick's law for the fluxes and write the conservation of the particles:

$$\vec{j}_i = -\nu X_i \vec{\nabla}.\mu_i \quad \text{and} \quad \frac{\partial X_i}{\partial t} + \vec{\nabla}.\vec{j}_i = 0 \qquad (8)$$

where $\nu$ is the mobility of the particles (identical for both populations) and $\mu_i$ are the chemical potentials of each population: $\mu_i = \partial f / \partial X_i$. In absence of hydrodynamic interactions, the only effect of one population onto the motion of the other arises from the term coupling the two concentrations in the free energy density (eq 7):

Combining eqs(7) and (8) in the conventional way and switching to the recipocal space, we obtain the kinetic equations for the two kinds of particles:

$$\frac{\partial \delta X_{q1}}{\partial t} = \delta X'_{q1} = -\nu q^2 \bar{X}_1 \left[\left(A p_1^2 + \frac{k_b T}{\bar{X}_1}\right)\delta X_{q1} + A \, p_1 p_2 \, \delta X_{q2}\right] \qquad (9a)$$

$$\frac{\partial \delta X_2}{\partial t} = \delta X'_{q2} = -\nu q^2 \bar{X}_2 \times \left[A \, p_1 p_2 \, \delta X_{q1} + \left(A p_2^2 + \frac{k_b T}{\bar{X}_2}\right)\delta X_{q2}\right] \qquad (9b)$$

where $q$ is the wave vector. Solving these coupled kinetic equations we easily get two very simple expressions for the characteristic relaxation times (see appendix 1 for the details):

$$\tau_f^{-1} = \nu q^2 k_b T \left[1 + \frac{A}{k_b T}(p_1^2 \overline{X}_1 + p_2^2 \overline{X}_2)\right] \tag{10a}$$

$$\tau_s^{-1} = \nu q^2 k_b T \tag{10b}$$

where the indices $s$ and $f$ stand for slow and fast respectively. As expected, both modes are diffusive. The fast mode explicitly depends on both the temperature and the enthalpic contribution arising from the repulsions between charges. Whereas the slow mode is totally independent of the enthalpic terms. This is precisely what we expected from the discussion in the previous section: the fast mode is the collective diffusion mode whereas the slow mode only depends on the temperature and can be identified with the self diffusion mode.

Of course this model is by far oversimplified. However, it captures most of the salient features of the real situation. As quoted in the previous section, an essential point in the real situation is that highly charged droplets are more repulsive than the others. This aspect is well taken into account by the choice we did for the enthalpic contribution $h$ taken as a function of the overall local density of charges; at constant total particle concentration ($X_1 + X_2$), regions richer in highly charged particles bring an extra positive contribution to the free energy. Following these views, we expect the self diffusion mode to disappear when the difference between the charges of the two populations vanishes. To check this point, we proceed one step further with the model and calculate the relative amplitudes of the two modes (we apply the equipartition theorem to the eigen combinations of $\delta X_1$ and $\delta X_2$ which are not energetically coupled in (7)). The tedious complete calculations are summarized in appendix 1. Unfortunately, the general expressions are prohibitively complicated. Nevertheless (see eq-A-9), the amplitudes of the two modes are q-independent consistently with the experimental results. Moreover, the amplitude of the slow diffusive mode is proportional to $(p_1 - p_2)$ (see

eq-A-9) and decreases to zero when the difference in the number of charges per droplet in the two populations vanishes.

By chance, the expressions for the amplitudes simplify considerably if we consider the case where both populations have the same concentrations: $\bar{X}_1 = \bar{X}_2 = \bar{X}/2$. This simplified case is relevant for the true situation where we expect the fluctuations of the number of charges per droplet above and below the average to be essentially symmetrical. Writing $p_1 = p + dp \quad p_2 = p - dp$, $\delta X_q = \delta X_{q1} + \delta X_{q2}$ and $A_k = A/k_bT$, we finally get for the autocorrelation function of the concentration fluctuations:

$$\langle \delta X(q,t) \delta X(q,0) \rangle = \frac{\bar{X}}{2 \times (p^2 + dp^2)} \times \left\{ \frac{2p^2}{\left(A\bar{X}(p^2+dp^2)+1\right)} \exp(-q^2 \, k_bT \, v \left(A\bar{X}(p^2+dp^2)+1\right) t) \atop + \quad 2dp^2 \quad \exp(-q^2 \, k_bT \, v \, t) \right\} \quad (11)$$

Interestingly, according to eq.11 the amplitude of the slow mode remains finite when A diminishes and equals zero. However, of course, (see eqs. 10, 11) the characteristic times of the two modes become identical in this limit so that the two modes can no longer be distinguished. This general evolution could be checked, for example, by tuning the electrostatic interaction and thus A by adding increasing amounts of salt.

In figure 8, the relative amplitudes of the modes are found almost independent on $\Phi$ and $\bar{p}$. This feature does not come out from eq 11 ( when $\bar{X}_1 = \bar{X}_2 = \bar{X}/2$). However we must keep in mind that we have not specified the functional dependence of the enthalpic term. From the expansion in eq 7, A is an unknown function of $p_1\bar{X}_1, p_2\bar{X}_2$ and in this context it is impossible to discuss this feature further.

## Conclusion

We report here on a second relaxation mode which to our knowledge has not mentioned in the earlier study of similar weakly charged microemulsions. In the paper of Gradzielski and Hoffmann[17] which dates back to 1994 it was experimentally impracticable to ascertain such a second relaxation mode . However the correlation functions in figure 13 of their paper display -with an increasing number of charges- the same trend as the one observed here and illustrated in figure 2. Evilevitch et al in their recent paper [18] pursue a different goal in their study of a weakly charged O/W microemulsion and displayed no correlation functions: they explore the influence of the electrostatic interaction introduced by the charges between microemulsion droplets on the properties measured in different experiments and compare it to theoretical predictions. They analyze the results of DLS in terms of the collective diffusion of the charged droplets.

To determine the origin of the two modes we considered successively two possibilities. The one put forth by Pusey et al [3,4] and by Weissman [33] where the self diffusion in charged colloidal suspensions is revealed , in DLS , by the polydispersity in size . And the one we proposed here where the self diffusion is revealed by the "polydispersity in the number of charge per droplet" . In the two descriptions , the fast relaxation mode is indeed identified with the collective diffusion of the droplets and we find a dependency of $D$ on $\bar{p}$ and $\Phi$ which closely parallels the one observed in[18] and the slow mode is identified with the self-diffusion of the droplets . The basic mechanism in the two descriptions is however different , in the first case size polydispersity induces difference in the scattering power of the droplets while in the second case polydispersity in charge induces different interactions between otherwise supposedly identical droplets . At this point no experimental evidence allows us to discard one or the other description , furthermore a mixture of both mechanisms may well be at work in most situations of charged and moderately polydisperse colloids .

## Acknowledgements.

The SANS experiments have been performed on line PACE at Laboratoire Léon Brillouin - CEA-CNRS. We thank Loic Auvray and Didier Lairez for their help . We are grateful to Martin In for fruitful discussions.

The authors acknowledge the referee for pointing out to them the possibility that polydispersity could be one of the origins of their experimental observations .

**Appendix 1**: Derivation of the characteristic times and relative amplitudes of the two relaxation modes.

The two charged species have respectively charges $p_1, p_2$ , average concentrations $c_1 \overline{X}$ and $c_2 \overline{X}$ with $c_1 + c_2 = 1$ and fluctuations of concentration $\delta X_1(t), \delta X_2(t)$ .Concentrations are in molar fraction. The overall concentration fluctuation, in reciprocal space, is $\delta X_q(t) = \delta X_{1q}(t) + \delta X_{2q}(t)$. Our objective is to calculate the autocorrelation function of the concentration fluctuations : $\quad C(q,t) = \left\langle \left( \delta X_q(t) \, \delta X_q(0) \right) \right\rangle$ (A-1)

$\delta X_{1q}(t)$ and $\delta X_{2q}(t)$ are the solutions of the coupled equations (eqs 9 ). They are solved using matrix methods:

$$\mathbf{B} = \begin{vmatrix} B_{11} = c_1 \left( (p_1^2 \, A_k) + \dfrac{1}{c_1 \, \overline{X}} \right) & B_{12} = c_1 \, p_1 \, p_2 \, A_k \\ B_{21} = c_2 \, p_1 \, p_2 \, A_k & B_{22} = c_2 \left( (p_2^2 \, A_k) + \dfrac{1}{c_2 \, \overline{X}} \right) \end{vmatrix} \quad \text{(A-2)}$$

where $A = k_b T \times A_k = \dfrac{\partial^2 h}{\partial (p X_p)^2}$ as explained in the text. And eqs 9 write:

$\delta \mathbf{X}'_q(t) = S \, \mathbf{B} \bullet \delta \mathbf{X}_q(t)$ with $S = -\nu \, q^2 \, k_b T \, \overline{X}$, $\delta \mathbf{X}_q(t) = (\delta X_{q1}(t), \delta X_{q2}(t))$ and $\delta \mathbf{X}'_q(t) = (\delta X'_{q1}(t), \delta X'_{q2}(t))$, the derivative of $\delta \mathbf{X}_q(t)$ with respect to time. It can be rewritten, in tems of the diagonalized matrix $\mathbf{L}$, and the eigenvectors $\delta \mathbf{X}_{dq}(t)$ :

$$\delta \mathbf{X}_{dq}(t) = S \, \Lambda \bullet \delta \mathbf{X}_{dq}(t) \tag{A-3}$$

The two eigenvalues of $\mathbf{B}$ are: $\quad \Lambda_{11} = A_k \, (c_1 \, p_1^2 + c_2 \, p_2^2) + \dfrac{1}{\overline{X}} \quad \Lambda_{22} = \dfrac{1}{\overline{X}}$

And the solutions of (A-3) write: $\quad \delta \mathbf{X}_{qd}(t) = \exp(S \, \Lambda \, t) \bullet \delta \mathbf{X}_{qd}(0) \tag{A-4}$

where, $\delta \mathbf{X}_{qd}(t) = \mathbf{P}^{-1} \bullet \delta \mathbf{X}_q(t)$ with $\mathbf{P}$ defined by : $\Lambda = \mathbf{P}^{-1} \bullet \mathbf{B} \bullet \mathbf{P}$ (A-4) can be written as: $\quad \delta \mathbf{X}_q(t) = \mathbf{P} \bullet \exp(S \, \Lambda \, t) \bullet \delta \mathbf{X}_{qd}(0) \tag{A-5}$

In eq A-5 $\vec{\delta X}_q(0)$ are the initial fluctuations, to calculate the autocorrelation function (eq A-1) we must perform an average over all possible thermal fluctuations and $\left\langle \left| \delta \mathbf{X}_q(0) \right|^2 \right\rangle$ are deduced from the equipartition theorem using the expression of the free energy variations due to the fluctuations. The variation of free energy upon fluctuations of the local concentrations of both species given by eq 7 is a quadratic form. We define the matrix $\mathbf{E}$:

$$\mathbf{E} = \begin{vmatrix} E_{11} = \left[ (p_1^2 A_k) + \dfrac{1}{c1 \overline{X}} \right] & E_{12} = p_1 \, p_2 \, A_k \\ E_{21} = E_{12} = p_1 \, p_2 \, A_k & E_{22} = \left[ (p_2^2 A_k) + \dfrac{1}{c2 \overline{X}} \right] \end{vmatrix}$$

$dF$ then writes: $\quad dF = \dfrac{kT}{2} \, {}^t\delta \mathbf{X} \bullet \mathbf{E} \bullet \delta \mathbf{X} = \dfrac{kT}{2} \, {}^t\delta \mathbf{X}_E \bullet \Gamma \bullet \delta \mathbf{X}_E \tag{A-6}$

$\delta \mathbf{X} = (\delta X_1, \delta X_2)$. The diagonal matrix $\Gamma$ is deduced, by standard procedures from $\mathbf{E}$, the eigenvalues are $\gamma_1, \gamma_2$ and the normalized eigenvectors are $\delta \mathbf{X}_E = (\delta X_{E1}, \delta X_{E2})$.

$\delta \mathbf{X} = \mathbf{Q} \bullet \delta \mathbf{X}_E$ with $\mathbf{Q}$ defined by $\Gamma = \mathbf{Q}^{-1} \bullet \mathbf{E} \bullet \mathbf{Q}$. Eq A-5 becomes :

$$\delta \mathbf{X}_q(t) = \mathbf{P} \bullet \exp(S \, \Lambda \, t) \bullet \mathbf{P}^{-1} \bullet \mathbf{Q} \bullet \delta \mathbf{X}_{Eq}(0) \tag{A-7}$$

After calculating the matrix $\mathbf{P}$, $\mathbf{P}^{-1}$ and $\mathbf{Q}$, we can develop eq A-7 and we obtain:

$$\delta X(q,t) = \delta X_1(q,t) + \delta X_2(q,t) = \frac{1}{(c_1 p_1^2 + c_2 p_2^2)\sqrt{E_{12}^2 + \Delta^2}} \times$$

$$\left\{ \begin{array}{l} [\exp(S\,\Lambda_{11}\,t)](c_1 p_1 + c_2 p_2)\,(\delta X_{1qE}(0)\,a1 + \delta X_{2qE}(0)\,b1) \\ + [\exp(S \times \Lambda_{22} \times t)](p_1 - p_2)\,(\delta X_{1qE}(0)\,a2 + \delta X_{2qE}(0)\,b2) \end{array} \right\}$$

(A-8)

with
$$\alpha = \frac{1}{2}(E_{11} + E_{22}) \qquad \beta = \frac{1}{2}(E_{22} - E_{11})$$
$$\delta = \sqrt{\beta^2 + E_{12}^2} \qquad \gamma_1 = \alpha + \delta \qquad \gamma_2 = \alpha - \delta$$
$$\Delta = (\gamma_1 - E_{11}) = -(\gamma_2 - E_{22}) = \beta + \delta$$
$$a_1 = (p_1 E_{12} + p_2 \Delta) \qquad b_1 = (p_2 E_{12} - p_1 \Delta)$$
$$a_2 = (c_1 p_1 \Delta - c_2 p_2 E_{12}) \qquad b_2 = (c_1 p_1 E_{12} + c_2 p_2 \Delta)$$

From eq A-8, rearranging terms, averaging and rejecting the cross product $\delta X_{q1E} \times \delta X_{q2E}$ which vanish upon averaging as the two fluctuations are statistically independent we obtain:

$$\langle \delta X(q,t) \delta X(q,0) \rangle = \frac{1}{\left((c_1 p_1^2 + c_2 p_2^2)\sqrt{E_{12}^2 + \Delta^2}\right)^2}$$

$$\left\{ \begin{array}{l} [\exp(-\tau_f^{-1} t)](c_1 p_1 + c_2 p_2)\left[\langle \delta X_{1qE}^2 \rangle a_1 (K1\,a_1 + K2\,a_2) + \langle \delta X_{2qE}^2 \rangle b_1 (K1\,b_1 + K2\,b_2)\right] \\ + [\exp(-\tau_s^{-1} t)](p_1 - p_2)\left[\langle \delta X_{1qE}^2 \rangle a_2 (K1\,a_1 + K2\,a_2) + \langle \delta X_{2qE}^2 \rangle b_2 (K1\,b_1 + K2\,b_2)\right] \end{array} \right\}$$

(A-9)

with $K1 = (c_1 p_1 + c_2 p_2)$ and $K2 = (p_1 - p_2)$

The values of $\langle \delta X_{1qE}^2 \rangle$ and $\langle \delta X_{2qE}^2 \rangle$ in the above formula are derived from the equipartition theorem applied to the two independent q-modes, in eq A-6:

$$\langle \delta X_{1qE}^2 \rangle = \gamma_1^{-1} \text{ and } \langle \delta X_{2qE}^2 \rangle = \gamma_2^{-1}$$

The relaxation times of the two modes are given by (eq 10):

$$\tau_f^{-1} = -S\,\Lambda_{11} = v\,q^2\,k_b T\,\bar{X}\left(A_k\,(c_1\,p_1^2 + c_2\,p_2^2) + \frac{1}{\bar{X}}\right)$$

$$\tau_s^{-1} = -S\,\Lambda_{22} = v\,q^2\,k_b T$$

Eq A-9 simplifies to eq 11 when when the two populations have equal concentration: $c_1 = c_2$